\documentclass[twocolumn]{aastex62}

\usepackage{color}
\usepackage{subfigure}
\usepackage{booktabs}
\usepackage{longtable}
\usepackage{bigstrut}

\revised{\today}

\shorttitle{MMA PTA Constraints on SMBHBs}
\shortauthors{Xin, Mingarelli, Hazboun}

\begin{document}

\title{Multimessenger pulsar timing array constraints on supermassive black hole binaries traced by periodic light curves }

\correspondingauthor{Chengcheng Xin}
\email{cx2204@columbia.edu}

\author[0000-0003-3106-8182]{Chengcheng Xin}
\affil{Columbia University, Department of Astronomy, 550 West 120th Street, New York, NY, 10027, USA}

\author[0000-0002-4307-1322]{Chiara M. F. Mingarelli}
\affil{Center for Computational Astrophysics, Flatiron Institute, 162 Fifth Ave, New York, NY, 10010, USA}
\affil{Department of Physics, University of Connecticut, 196 Auditorium Road, U-3046, Storrs, CT 06269-3046, USA}

\author[0000-0003-2742-3321]{Jeffrey S. Hazboun}
\affil{University of Washington Bothell, 18115 Campus Way NE, Bothell, WA 98011, USA}

\begin{abstract}
Supermassive black hole binary systems (SMBHBs) emitting gravitational waves may be traced by periodic light curves. We assembled a catalog of 149 such periodic light curves, and using their masses, distances, and periods, predicted the gravitational-wave strain and detectability of each binary candidate using all-sky detection maps. We found that the International Pulsar Timing Array (IPTA) provides almost uniform sky coverage -- a unique ability of the IPTA -- and by 2025 will improve NANOGrav's current minimum detectable strain by a factor of 6, and its volume by a factor of 216. Moreover, IPTA will reach detection sensitivities for three candidates by 2025, and 13 by the end of the decade, enabling us to constrain the underlying empirical relations used to estimate SMBH masses. We find that we can in fact already constrain the mass of a binary in Mrk 504 to $M<3.3\times 10^9~M_\odot$. We also identify 24 high-mass high-redshift galaxies which, according to our models, should not be able to host SMBHBs. Importantly the GW detection of even one of these candidates would be an essentially eternal multimessenger system, and identifying common false positive signals from non-detections will be useful to filter the data from future large-scale surveys such as LSST.
\end{abstract}


\section{Introduction} 

Gravitational waves (GWs) emitted by supermassive black hole binary (SMBHB) systems with periods of years to decades are expected to be the most powerful sources of GWs in the Universe. Searches for individual SMBHBs can be aided by looking for periodic light curves in Active Galactic Nuclei (AGN), which trace the SMBHB activity and serve as electromagnetic (EM) counterparts to these GW sources (e.g. \citealt{KauffmannEtAl:2000, GouldingEtAl:2018}).

With decades of precision pulsar timing data, Pulsar Timing Arrays (PTAs) are ready and able to make the first detection of low-frequency GWs \citep{SBS:2019}. Ultra-stable radio millisecond pulsars are the best clocks in nature -- deviations in their arrival times manifesting over years to decades can therefore signal the presence of passing low-frequency GWs \citep{hd83}.

 Identifying the host galaxy of a SMBHB system would yield an essentially eternal multimessenger system --- the black holes are millions of years from coalescence, and would provide a treasure trove of science: direct evidence that the infamous final parsec problem is solved \citep{BegelmanBlandfordRees:1980}, clues as to how it was solved by the quantity of gas, stars, and residual eccentricity in the system \citep{MM01, Sesana:2013}, as well as probe General Relativity, e.g. \citet{OBeirneEtAl:2019}.

  Here we assembled a catalog of 149 periodic light curves which may trace SMBHB activity, including 111 SMBHB candidates from Catalina Real Time Transient Survey ({\it CRTS}; \citealt{Graham2015b}), 33 from Palomar Transient Factory ({\it PTF}; \citealt{Charisi2016}), 3 from {\it Pan-STARRS} \citep{Liu2019}, OJ287 \citep{1996ApJ...460..207L} and 3C66B \citep{Sudou2003,Arzoumanian2020}, 
  though in the end we restrict our analysis to the \cite{Graham2015b} sample since we found that the other candidate binaries are too high-frequency to be detected.  

  Using published binary parameters, we predict the GW strain of each binary candidate, and for the first time predict their time to detection by constructing all-sky detection maps to simulate the IPTA in 2025 (IPTA2025), Phase 1 of the Square Kilometre Array (SKA1), and its second phase (SKA2) using {\tt hasasia} \citep{HazbounPRD:2019}, from hereon H19.

  These public data, simulations, and open-source codes will help to inform where to search for new pulsars to accelerate GW detection, and lead to targeted GW searches -- making these searches more sensitive by at least factor of 2 \citep{Arzoumanian2020}, more if the GW frequency is known.

\section{Gravitational-Wave Signals} \label{sec:strains}
 Previously, the periodic light curves from \citet{Graham2015b}, from hereon G15, were used by \citet{SesanaEtAl:2018} to test the binary hypothesis: they showed that if all the AGN light curves were really SMBHBs, that the inferred cosmic population of SMBHBs would create a GW background currently in tension with PTA upper limit at the time~\citep{NG11yr_gwb}. \citet{KelleyEtAl:2019} also carried out simulations which predict that, statistically, five of the {\it CRTS} sources may contain genuine binaries.

We now have new tools which we can apply directly to the {\it CRTS} sample for the first time: {\tt hasasia} -- a public code which can create continuous GW (CGW) detection sky maps, and new results from \citet{Arzoumanian2020} which show that targeted searches can boost CGW sensitivity by at least a factor of two.  Indeed, it is only now possible to rigorously investigate the detectability of each individual binary candidate's GW signal.

The GW strain amplitude for a SMBHB is determined via
\begin{equation}
\label{eq:strain}
    h = \frac{2\mathcal{M}_c^{5/3}(\pi f)^{2/3}}{D_L} \, ,
\end{equation}
where $\mathcal{M}_c^{5/3} = q/(1+q)^2 M^{5/3}$ is the chirp mass, $M=M_1+M_2$ is the total binary mass, $q=M_2/M_1\leq 1$ is the mass ratio, $D_L$ is the luminosity distance to the source and $f$ is the observed GW frequency, and we let $G=c=1$. Many of the mean BH mass estimates are reported in previous literature that identified periodic sources, e.g. G15, \cite{Charisi2016, Liu2019}. While G15 computed the strain in their Figure 10, they did not include errors on the total mass $M$ (their error bars in Figure 10 are from varying $0.05\leq q \leq 1.0$).
This mass error is crucial for estimating the error on the strain $h$, since $h\propto M^{5/3}$, and is therefore even more important than the error on the mass ratio $q$.

We compute the SMBH masses as in G15, using the widths of H$\beta$ and Mg-II spectral emissions \citep{Shen2008}. We redo this calculation directly from SDSS spectra\footnote{We extract interactive spectra from SDSS Data Release 13 (DR13); \url{https://www.sdss.org/dr13/spectro/} \citep{Albareti2017}.}
since G15 do not report mass errors which are crucial for determining the uncertainties in $h$. In order to get a handle on these errors, we generate $160,000$ Monte Carlo (MC) samples to create a distribution of the mass, $M$, and also sample in the ratio $q$ from uniform distributions over the range 0.25 to a maximum of 1. The lower limit of $q$ is somewhat arbitrary, but since the BH mass error dominates the uncertainty in $h$, the choice of the lower limit on $q$ is of little consequence, as smaller values of $q$ simply extend the lower value of the error bar on $h$. It is also interesting to note that while the relativistic Doppler boost model is very important in some binary SMBH systems with small mass ratios, e.g.  \cite{Dorazio2015Nature,Charisi2018}, this subcategory of SMBHB is difficult to detected due to this small $q$ value.  

We interpret the peak of the distribution as the mean, $h_0$ and the error bars are taken as one standard deviation ($1\sigma$) in the mass distribution. While only 36 {\it CRTS} sources have the available SDSS spectra needed for this calculation, this method is generic and can be applied to a larger sample when spectra are available. We report all our mass calculations and their error estimates in \autoref{Table:BHmass_err}.

The GW frequencies are obtained via reported binary orbital periods $P$, i.e. $f _{GW}= 2/P$. The luminosity distance $D_L$ is computed via the redshift of the source, assuming cosmological parameters, $H_0=0.7 $ km Mpc$^{-1}$ s$^{-1}$ and $\Omega_m=0.3$. 

\begin{figure*}[ht!]
  \includegraphics[width=\columnwidth]{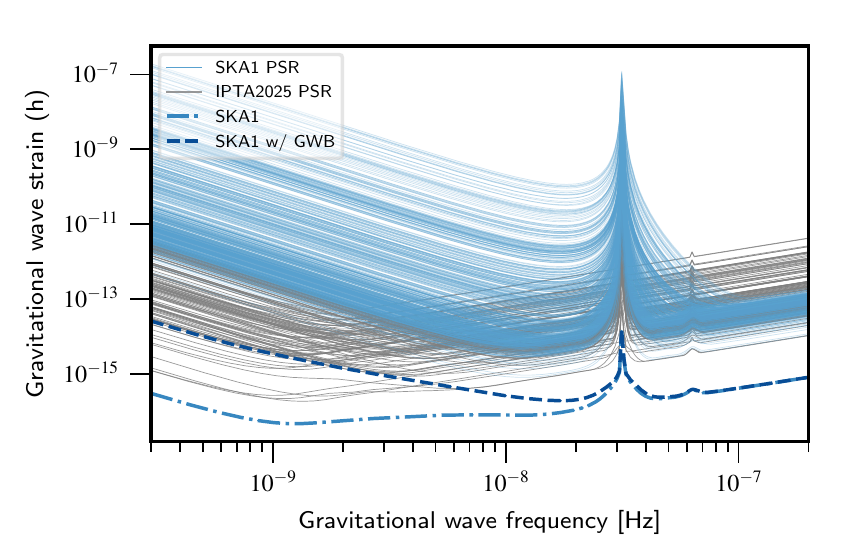}
 \includegraphics[width=\columnwidth]{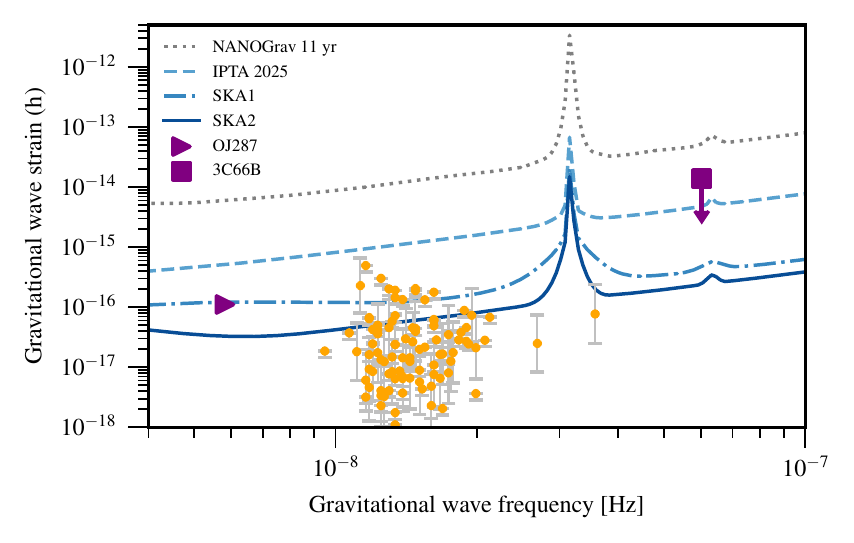}
 \caption{{\bf Current and Future GW Detection Curves}. Left: Individual SKA1 pulsar sensitivities and full PTA sensitivity curves. Pulsars with short observing time spans considerably increase the sensitivity at higher frequencies. The atypical flat bucket of the curve comes from the addition of low RMS - short time span pulsars to IPTA pulsars. An unresolved  GW background with $A_{\rm GWB}=1.92\times10^{-15}$ deteriorates our sensitivity to individual SMBHBs by at most a factor $\sim 3$ for the lowest frequency {\it CRTS} sources. Right: Truncated GW frequency regime relevant to {\it CRTS} sources (orange circles with $1-\sigma$ error bars), OJ287, and 3C66B. The continuous curves are the S/N $=3$ detection curves for NANOGrav, IPTA and SKA1 and 2 experiments. We only plot candidates with strain $h>10^{-18}$, GW strain less than this is unlikely to be detectable by PTAs.}
 \label{Fig:P_CRTS}
\end{figure*}

An important consideration for the detection of these individual SMBHB systems is the existence of a loud GW background (really a foreground), likely generated by the cosmic merger history of SMBHs. Currently it is unknown what the true amplitude of this GW background is, though the most recent estimate is $A\sim2\times10^{-15}$ \citep{NG_gwb_12p5}. Potential individual GW amplitudes in the datasets we investigate here are much lower than this, and at best, at the level of some of the most conservative GW background models \citep{Ryu2018,Bonetti2018, SesanaEtAl:2018, ZhuEtAl:2019}. We therefore study the impact of this stochastic background signal on the detection of the individual SMBHB signals. On the left-hand side of \autoref{Fig:P_CRTS}, we illustrate the effect of a stochastic GW background on the sensitivity to single sources as a separate noise source, though the background is obviously tied in with individual sources, especially at the detection threshold. The number of resolvable binaries in a single frequency bin -- the so-called confusion limit for PTAs -- varies in the literature from 4 \citep{Babak2012} to $2N-7$ \citep{Boyle2012}, where $N$ is the number of pulsars in the PTA. Ever evolving data analysis techniques \citep{Becsy2020} will allow individual sources to be pulled from the GWB as they become significant in more sensitive datasets. On the right-hand side of \autoref{Fig:P_CRTS}, we show the strain amplitudes (orange dots) given by eq.~\ref{eq:strain} and the errors of $h$ for the {\it CRTS} candidates, OJ287 and 3C66B, along with the detection curves for NANOGrav, IPTA, SKA1 and SKA2 with S/N = 3. The values of $h$ can be compared with Fig. 6 in \citet{SesanaEtAl:2018}, where they show the top $\sim$half of the {\it CRTS} candidates with the strongest GW signals. In \S~\ref{sec:results} we compare our top candidates with those in \citet{SesanaEtAl:2018}.

Since our catalog of candidate SMBHBs are mid-to-high GW frequency for PTAs, we also compute their GW frequency evolution, $\dot{f}$,
\begin{equation}
    \label{eq:fdot}
    \dot{f}= \frac{96}{5} \pi^{8/3} \mathcal{M}_c^{5/3}{f}^{11/3}  \, .
\end{equation}
The expert reader will notice that small values of $\dot{f}$ mean that the Earth term and the pulsar term lie in the same GW frequency bin, and will thus double a pulsar's residual. This has been taken into account in {\tt hasasia}, H19.

\section{Electromagnetic counterparts}
\label{sec:EM}
Hydrodynamical simulations have demonstrated that bright quasar variability can be modulated periodically by the orbital period of the binary SMBH, due to perturbations by the surrounding gas in the circumbinary accretion disk, see e.g. \citealt{Roedig2014,Farris+2014,Dorazio+2016} and references therein.

A promising observational approach to identify SMBHBs in the optical bands is to search for quasars with periodic variability.
One of the first candidate SMBHBs, OJ287, was identified with variable observed luminosity fluctuations in the form of repeating outbursts occurring every $\sim$12 years \citep{1996ApJ...460..207L, LaineEtAl:2020}. Another prominent object is an unequal-mass SMBHB candidate in quasar PG1302-102. It emerged from a systematic search in {\it CRTS} for its quasi-sinusoidal optical variability \citep{Graham2015a} possibly due to the fact that the emissions from its mini-disc are Doppler boosted \citep{Dorazio2015Nature,Xin2019}. 

Infrared (IR) variability can also be caused by a SMBHB heating its surrounding dust torus, see e.g. \citealt{Antonucci1993, Krolik1988}. In fact, \cite{JunEtAl2015} and \cite{DOrazio2017} implemented the IR reverberation modelling to a Doppler boosted system, PG1302-102, and reproduced its periodicity in the optical lightcurve in IR wavelength. As such, candidate binaries may further benefit from follow-up measurements by the James Webb Space Telescope, {\it JWST}.  

In addition to periodicity, spectral signatures across optical, UV and X-ray bands are widely used to differentiate binary SMBHs from normal AGNs powered by single SMBHs. 
Merging binary SMBHs likely have different X-ray spectral profiles from isolated SMBHs, including harder X-ray spectra \citep{Roedig2014,Farris2014b,Ryan2017}.
Of particular relevance is \citealt{Saade2020}, who measured X-ray spectra of 7 {\it CRTS} candidates within the {\it Chandra} X-ray energy range and computed their optical/UV-to-X-ray spectral indices. While their spectra showed no difference from the broader AGN population with a purported single SMBH, they are careful to note this is not entirely unexpected: in some theoretical models (e.g., \citealt{Roedig2014}), thermal X-ray profiles can only distinguish binaries separated by 100$r_g$, where $r_g = GM/c^2$ is the gravitational radius.

Interestingly, \citealt{Saade2020} identified HS 1630+2355 (also known as FBQS J163302.6+234928) as the only AGN in their sample that could host a SMBHB since its semi-major axis is $a\sim 57r_g$, making it a tantalizing SMBHB candidate, which we highlight moving forward.

\section{Forecasting PTA capabilities}
In addition to cataloging current candidate SMBHB systems, we estimate when these binaries may be detected.
Using H19's open source Python package {\tt hasasia}, we assess the current sensitivity of NANOGrav and forecast the sensitivity of future PTA experiments. For resolvable individual signals from SMBHBs this involves using a matched filter statistic and building an effective strain-noise power spectral density (interpreted as the sensitivity) using the sky location, detector response functions, and noise parameters of the pulsars. The detection thresholds are then calculated using the expectation value of the S/N for a circular binary given the sensitivity. There are many subtleties involved with calculating a PTA's sensitivity to various GW sources, and the interested reader in encouraged to refer to H19 for details necessarily left out here.

There are a number of statistics in the literature \citep{Babak2012, Ellis2012, Taylor2014} developed for single-source (CGW) searches in PTA data sets. Here we use the S/N from \citet{HazbounPRD:2019}, as it was developed with calculations of generic sky sensitivities in mind. In our forecasts we have used the full RMS errors quoted by current PTA data release papers, but have avoided explicitly injecting any time-correlated (red) noise into these pulsars, since it is unclear how much of the red noise currently observed in PTA pulsars is due to the stochastic background of GWs and how much is due to intrinsic spin noise. It is also likely that in the SKA era, if red noise models are not sufficient for PTA use, then it will be feasible to avoid the small number of millisecond pulsars \citep{NG11yr_data} with large red noise amplitudes.

The NANOGrav 11-year detection curve is based on the noise parameters in \citet{NG11yr_data}. The IPTA2025 curve uses the pulsars and noise parameters from \citet{PereraEtAl:2019} as a starting place, and builds the array by adding 4 pulsars per year with RMS values of $300\pm 100$~ns. In order to add 4 IPTA pulsars per year to those in \citet{PereraEtAl:2019} we use the sky positions of the current pulsars to build empirical distributions for drawing new pulsar positions. The RMS values for the pulsars are drawn from a truncated normal distribution with mean and standard deviation of $300\;{\rm ns}\pm100\;{\rm ns}$, truncated at $10\;{\rm ns}$ and $600\;{\rm ns}$. The cadences are pulled from an empirical distribution based on those in \citet{NG11yr_data,PereraEtAl:2019}, where any new pulsar with an RMS less than $300\;{\rm ns}$ is observed weekly. 

The SKA sensitivity estimates are based on projected IPTA data with a large addition of pulsars distributed according to the planned SKA MID and LOW surveys in the first few years of the SKA1 \citep{KeaneEtAl:2015}. Both conservative and more optimistic estimates for SKA1 and 2 are included here.
The SKA LOW survey will concentrate on regions further than $\pm5^\circ$ from the galactic plane, while the SKA MID survey will concentrate on regions within $10^\circ$ of the galactic plane \citep{KeaneEtAl:2015}. We assume that $15\%$ of the millisecond pulsars discovered will be suitable for PTAs, meaning that SKA1 will have $\sim 675$~millisecond pulsars, and SKA2 will have $120$ more. All pulsars in IPTA2025 are used in the SKA with extended baselines from IPTA DR2 \citep{PereraEtAl:2019}, as it is assumed that any SKA PTA will be based in large part on the extensive data sets from other PTAs. 

We offer both optimistic and conservative outlooks for constraining SMBHB candidate masses in \autoref{Table:table_of_strains} (detection prospects in \autoref{Table:table_h0}). Our conservative sensitivity projections use the same empirical distribution of RMS errors as are used to build IPTA2025, while the optimistic sensitivities are built using a distribution more inline with the pulsars' RMS being ``jitter limited'' \citep{Shannon_Jit:2014,LamShortTime:2016,LamJitter:2019}, with mean and standard deviation of $100\;{\rm ns}\pm30\;{\rm ns}$, truncated at $9\;{\rm ns}$ and $201\;{\rm ns}$. In general we leave the RMS values of the adopted IPTA pulsars unchanged, except for PSR J1640+2224 which is extremely close in sky position to HS $1630+2355$, and the two best timers in this region of the sky, PSR J1909-3744 and PSR J1713+0747. These three pulsars are assumed to be observed by the SKA and their RMS values for the time spans considered here are set to the jitter values from \citet{LamJitter:2019}.

All PTA detection curves are presented as S/N $=3$ thresholds on the GW strain\footnote{There is a factor of $2$ difference in the definition of $h_0$ between H19 and this manuscript which has been taken into account for all of our calculations.}. This is the threshold set by the PTA community for detection of single sources in the nanohertz band. These forecasts represent only a limited number of the many permutations one could postulate, hence the Jupyter Notebooks used to make these projections are available on \href{https://github.com/Hazboun6/pta_forecasts}{GitHub}.

\subsection*{Special considerations for multimessenger signals}
When a periodic light curve or another EM tracer for a SMBHB system is identified, there are some concrete steps to take to increase a PTA's sensitivity to the candidate GW source.

It is our good fortune that the top two potential SMBHB host galaxies in \autoref{Table:table_of_strains} lie either in, or very close to, both the European PTA (EPTA) and NANOGrav's most sensitive region of the sky~\citep{Babak2016, NG11yr_cgw}. This already improves their detection prospects. Moreover, if a pulsar is close to the sky location of a potential SMBHB system, one also gains a factor of a few in sensitivity from the pulsar detector response function. We can take concrete steps to increase the S/N ($\rho$) of a potential GW candidate:
$\rho \propto \left<NTc/\sigma^2\right>^{1/2}$, where for simplicity the N pulsars have the same intrinsic properties, $T$ is the length of the dataset, $c$ is the cadence of the observation, and $\sigma$ is the white noise RMS, e.g. 
\citealt{NG11yr_data}. 
We can therefore increase the S/N by increasing the number of pulsars, the pulsar observing cadence $c$, and spending more time on observing these in an effort to decrease the RMS white noise value $\sigma$~\citep{BurtEtAl:2011, Mingarelli2017, Lam_opt:2019}. Moreover combining pulsar datasets under the auspices of the IPTA is an excellent way to readily increase a pulsar's total observing time $T$ and cadence $c$.

Furthermore, \citealt{Arzoumanian2020} show that targeted GW searches increase NANOGrav's GW sensitivity by at least a factor of two -- more if there is GW frequency information.

Some things are left to chance: both NANOGrav and EPTA show uneven CGW sensitivity on the sky due to the largely anisotropic distribution of pulsars, and their uneven timing properties. If we are fortunate, the source will lie in an area of high sensitivity. Moreover, the antenna beam pattern (or detector response function) of a pulsar in direction $\hat p$ to a GW is $ \propto 1/(1+\hat\Omega\cdot\hat p)$, where $\hat\Omega$ is the direction of GW propagation. This response function has been well-studied, and is clearly maximal for CGW sources in direction $-\hat{\Omega}$, since the denominator becomes small as $-\hat\Omega \rightarrow \hat{p}$. At exactly $\hat{\Omega}\cdot \hat{p}=-1$ the response is zero, due to surfing effects \citep{Baskaran:2008za,  CX12, MS14}. If we are very fortunate, the GW source will not only lie in a sensitive sky region, but also be closely aligned with a pulsar.

\section{Unlikely Binaries in {\it CRTS}} \label{sec:probability}
\citet{Mingarelli2017} computed the probability of a galaxy hosting a SMBHB system emitting nanohertz GWs:
\begin{equation}
\label{eq:pmerge}
    P_{\rm merger} = \frac{t_c}{T_z}\int_{0.25}^1 d\mu_\ast \frac{dN}{dz}(M_{\ast},z,\mu_\ast)T_z \,
\end{equation}
where $t_c = 5/256 (\pi f)^{-8/3} \mathcal{M}_c^{-5/3}$, $dN/dz$ is the cumulative galaxy-galaxy merger rate \citep{Rodriguez-Gomez2015}, $\mu_*$ is the stellar mass ratio of the parent galaxies, and  $T_z$ is the estimated binary lifetime.

Briefly, this probability is the product of the probability of two factors: the existence of a pair of SMBHs in the galaxy resulting from a galaxy merger (the integrand in \autoref{eq:pmerge}), and the probability that this binary is emitting nanohertz GWs, $t_c/T_z$. Here the binaries overcome the final parsec problem \citep{BegelmanBlandfordRees:1980} by stellar hardening \citep{Quinlan96}.

We use the public software Nanohertz GWs \citep{MingarelliGWs} to compute this probability of all the AGN in our sample. 
Since the AGN light curve catalog we assembled is not complete, we can only provide a relative ranking of the probability of a given AGN hosting a SMBHB.

Systems at higher redshift may contain more gas, hence the SMBHBs may overcome the final parsec problem by gas interactions in addition to stellar hardening, e.g. \citet{Sesana:2013, TiedeEtAl:2020}. The gas and stellar 3-body interactions may also lead to large binary eccentricities, which is a subject of future study and could also help to solve the final parsec problem. A follow-up paper will include more such solutions to the final parsec problem.

\section{Results} \label{sec:results}

Of the 111 {\it CRTS} sources, mass estimates are available for 98. We therefore compute the strain for 136 SMBHB candidates identified in {\it CRTS}, {\it PTF}, {\it Pan-STARRS1} surveys, alongside individual quasars OJ287 and 3C66B, and compare the strains with current and future PTA detection curves. None of the {\it PTF} and {\it Pan-STARRS1} candidates are detectable by SKA2, because these candidates generally have larger masses and very low GW frequencies to be detectable by any PTA -- we therefore focus our attention on {\it CRTS}. 

A multimessenger signal could be an important test of GR by comparing the change in the observed EM period and the change in the GW frequency, akin to the Hulse-Taylor binary \citep{ht75}. As such, we compute the GW frequency derivative, $\dot{f}_{\mathrm{gw}}$, \autoref{eq:fdot}, for all the SMBHB candidates. We find that most of the {\it CRTS} candidates would have a GW frequency shift of $\sim 10^{-4}$~{nHz/yr}, and would therefore not be detectable. We report the list of computed $\dot{f}$ values in \autoref{Table:BHmass_err}. However, we find that for 3C66B $\dot{f}_{\mathrm{gw}}= 0.14$~nHz/yr. Since \citet{Sudou2003}'s claim that 3C66B is a binary with an orbital period of $1.05 \pm 0.03 $~years, the source should have evolved from 60.4~nHz to 62.8 nHz over the last 17 years if it were indeed a SMBHB system. Today, it would have a period of $P=2/62.8$~nHz = 1.01 yrs -- an evolution of 15 days since the initial measurement. The error on the original period is about 11 days, so new EM measurements may prove to be illuminating.

Importantly galaxy 3C66B's detection prospects leap by almost an order of magnitude from the 11-year data constraints \citep{Arzoumanian2020} to IPTA2025, \autoref{Table:table_h0}. Much of this increase in detectability is due to the IPTA's enhanced sky coverage, even over IPTA DR2. Similar to the estimates for PSR J1640+2224, PSR J0218+4232's proximity to 3C66B makes it an ideal target for longer observing periods, and a high-cadence timing campaign.

We also find that we will be able to constrain SMBHB masses earlier than we can detect them, so we report two strain values: $h_{1\sigma}$ which is the $1-\sigma$ upper limit value of the strain, \autoref{Table:table_of_strains}, and $h_0$, which is the maximum a posteriori value of the strain, \autoref{Table:table_h0}. When we refer to constraints on the mass we refer to $h_{1\sigma}$ -- this is the point where we can start to constrain the mass upper limits -- and for detection claims we refer to $h_0$. While the mass ratio $q$ will be uncertain, these upper bounds on strain are dominated by total mass uncertainty.

\begin{figure*}[ht!]
 \includegraphics[width=2\columnwidth]{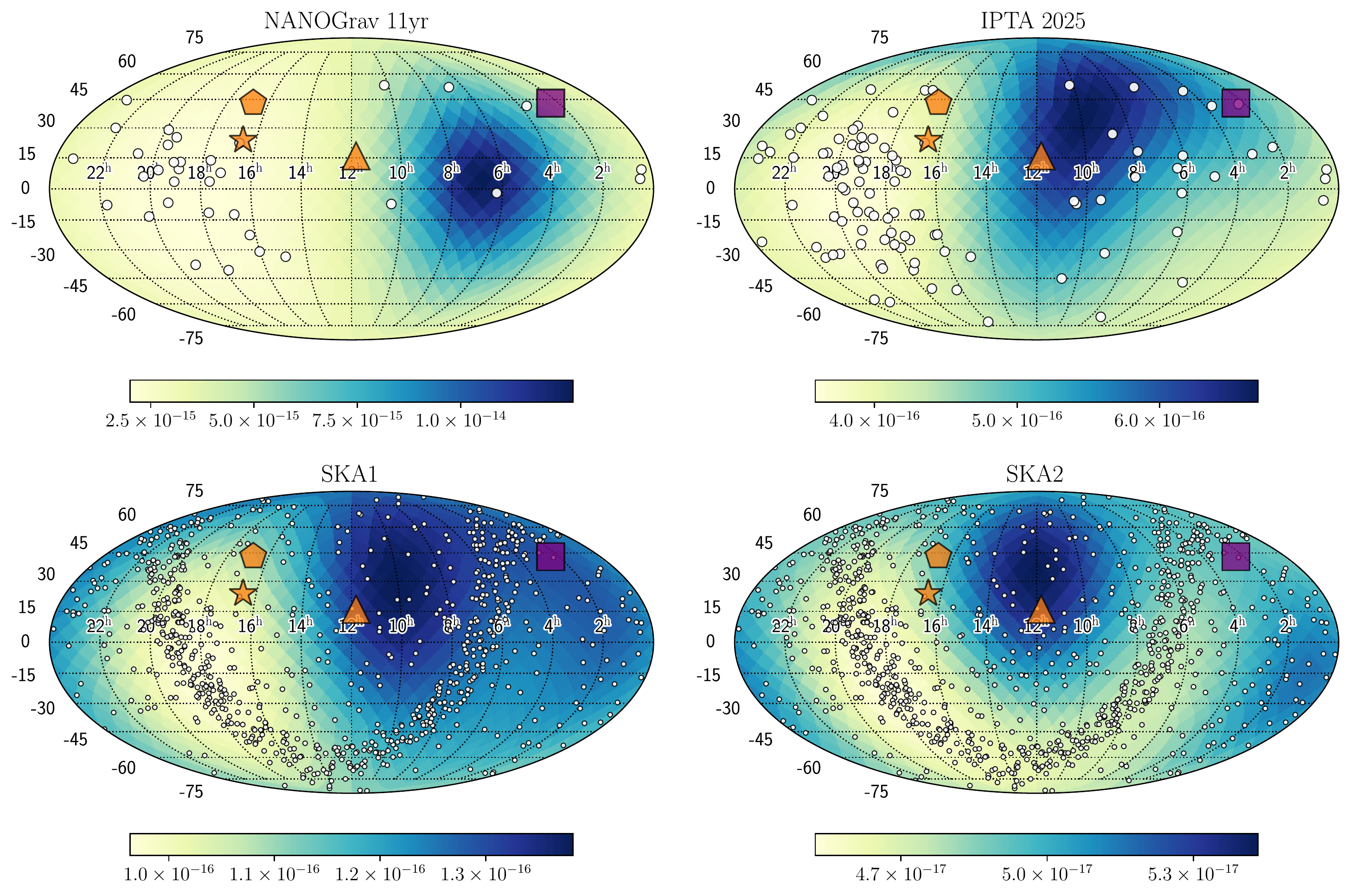}
 \caption{{\bf Minimum strain $h$ for an S/N $=3$ detection for current and future PTAs}. 
 {\it CRTS} GW candidates are marked in orange: HS 1630+2355 (orange star), SDSS J164452.71+4307 (orange pentagon) and SDSS J114857.33+1600 (orange triangle) are shown over the detection threshold of S/N $=3$ for the NANOGrav 11-year data set, and the {\tt hasasia}-modeled IPTA2025, SKA1 and SKA2 PTAs at $11.8 {\rm nHz}$. The purple square is 3C66B, possessing the largest S/N in \autoref{Table:table_of_strains}. While not in this frequency bin, the relative differences and proximity to pulsars is a useful comparison. Pulsars are small white circles, with their locations determined by planned SKA LOW and MID surveys \citep{KeaneEtAl:2015}. }
 \label{Fig:strain_sky_ptas}
\end{figure*}

\begin{table*}
\tablenum{1}
\caption{Top 13 periodic SMBHB candidates ranked by strain upper limit (UL; $h_{1\sigma}$), where the mass can be constrained by SKA2 (11 in {\it CRTS} plus OJ287 and 3C66B). For detection claims we use $h_0$, \autoref{Table:table_h0}. A further 15 candidates have marginal S/N $\sim 3$, but are not shown here. The last four columns report the S/N ($\rho$) on the $1-\sigma$ strain value for current and future PTA experiments. In the last two columns the S/N values are listed as \textit{optimistic(conservative)}. Importantly, \citet{Arzoumanian2020} find that strain ULs improve by a factor of at least two in targeted GW searches, which we do not take into account here. *For 3C66B $h_{1\sigma}$ is the 95\% UL from \citet{Arzoumanian2020}. \label{Table:table_of_strains}}
    \centering
    \begin{tabular}{lllccccccc}
\hline \hline    
Object Name   & RA      & Dec      & $f_{GW}$ & log($M$) & strain & NG $\rho$, & IPTA $\rho$, & SKA1 $\rho$, & SKA2 $\rho$, \\ 
& & & [Hz] &[$M_{\odot}$] & $h_{1\sigma}$ &  11-yr & 2025 & 2030 & 2034+\\
\hline 
3C66B*  & 02 23 11.5 & +42 59 30   & 6.04E-08  & 9.08  & 1.40E-14   & 1.60 & 15.7 & 82.2(35.1) &164(69.5)  \\
HS 1630+2355         & 16 33 02.7 & +23 49 28.8 & 1.13E-08 & 9.86   & 6.25E-16 & 0.9      & 4.86         & 18.32(8.19) & 40.45(16.51) \\
SDSS J164452.71+4307 & 16 44 52.7 & +43 07 52.9 & 1.16E-08 & 10.15  & 4.94E-16 & 0.65     & 3.67         & 13.77(6.26) & 30.82(12.73) \\
SDSS J114857.33+1600 & 11 48 57.4 & +16 00 22.7 & 1.25E-08 & 9.9    & 3.03E-16 & 0.22     & 1.39         & 6.88(2.74)  & 16(6.27)     \\
HS 0926+3608         & 09 29 52.1 & +35 54 49.6 & 1.48E-08 & 9.95   & 2.04E-16 & 0.08     & 0.74         & 4.29(1.66)  & 9.52(3.68)   \\
SDSS J092911.35+2037 & 09 29 11.3 & +20 37 09.2 & 1.30E-08 & 9.92   & 2.02E-16 & 0.07     & 0.87         & 4.43(1.73)  & 10.89(4.17)  \\
SDSS J133516.17+1833 & 13 35 16.1 & +18 33 41.8 & 1.34E-08 & 9.76   & 1.91E-16 & 0.18     & 1.00            & 4.67(1.92)  & 9.7(3.87)    \\
SDSS J140704.43+2735 & 14 07 04.5 & +27 35 56.3 & 1.48E-08 & 9.94   & 1.89E-16 & 0.17     & 0.95         & 4.50(1.86)   & 8.90(3.56)    \\
SDSS J134855.27-0321 & 13 48 55.3 & --3 21 41.4 & 1.62E-08 & 9.89   & 1.78E-16 & 0.14     & 0.8          & 4.09(1.64)  & 7.95(3.17)   \\
SDSS J160730.33+1449 & 16 07 30.3 & +14 49 04.2 & 1.34E-08 & 9.82   & 1.45E-16 & 0.17     & 0.95         & 3.97(1.71)  & 7.94(3.22)   \\
SDSS J131706.19+2714 & 13 17 06.2 & +27 14 16.7 & 1.39E-08 & 9.92   & 1.34E-16 & 0.11     & 0.63         & 3.08(1.26)  & 6.50(2.57)    \\
SNU J13120+0641      & 13 12 04.7 & +06 41 07.6 & 1.55E-08 & 9.14   & 1.33E-16 & 0.1      & 0.57         & 2.98(1.19)  & 5.97(2.37)   \\
OJ287   & 08 54 48.9 & +20 06 31  & 5.82E-09 &  10.26  & 1.11E-16   & 0.06 & 1.05 & 1.92(1.33) & 9.69(3.69)\\
\hline
\end{tabular}
\end{table*}

\begin{table*}
\tablenum{2}
\caption{Top 13 periodic SMBHB candidates, ranked by the mean strain ($h_0$),  which can be detected or constrained by SKA2 (11 in {\it CRTS} plus OJ287 and 3C66B). A further 4 candidates have marginal S/N $\sim 3$, but are not shown here. The last four columns report the S/N ($\rho$) on $h_0$ -- the maximum a posteriori strain value --  for current and future PTA experiments. The last two columns list the \textit{optimistic(conservative)} S/N values. The S/N calculations here do not include the additional factor of two one achieves from a targeted search \citep{Arzoumanian2020}. *For 3C66B $h_0$ is calculated in \citet{Arzoumanian2020}. \label{Table:table_h0} }
    \centering
    \begin{tabular}{lllccccccc}
\hline \hline
Object Name   & RA      & Dec      & $f_{GW}$ & log($M$) & strain & NG $\rho$, & IPTA $\rho$, & SKA1 $\rho$, & SKA2 $\rho$, \\ 
& & & [Hz] &[$M_{\odot}$] & $h_0$ &  11-yr & 2025 & 2030 & 2034+\\
\hline 
3C66B*  & 02 23 11.5 & +42 59 30   & 6.04E-08  & 9.08  & 7.2E-15  & 0.82  & 8.06 & 42.27 (18.06) &  84.59 (35.72) \\   
HS 1630+2355  & 16 33 02.7 & +23 49 28.8 & 1.13E-08 & 9.86   & 2.29E-16 & 0.33     & 1.78         & 6.71(3.00)      & 14.82(6.05)  \\
SDSS J164452.71+4307 & 16 44 52.7 & +43 07 52.9 & 1.16E-08 & 10.15  & 4.94E-16 & 0.65     & 3.67    & 13.77(6.26)  & 30.82(12.73) \\
SDSS J114857.33+1600 & 11 48 57.4 & +16 00 22.7 & 1.25E-08 & 9.9    & 3.02E-16 & 0.22     & 1.39       & 6.86(2.73)   & 15.95(6.25)  \\
HS 0926+3608         & 09 29 52.1 & +35 54 49.6 & 1.48E-08 & 9.95   & 2.04E-16 & 0.08     & 0.74       & 4.29(1.66)   & 9.52(3.68)   \\
SDSS J092911.35+2037 & 09 29 11.3 & +20 37 09.2 & 1.30E-08 & 9.92   & 2.02E-16 & 0.07     & 0.87       & 4.43(1.73)   & 10.89(4.17)  \\
SDSS J133516.17+1833 & 13 35 16.1 & +18 33 41.8 & 1.34E-08 & 9.76   & 1.91E-16 & 0.18     & 1.00      & 4.67(1.92)   & 9.7(3.87)    \\
SDSS J140704.43+2735 & 14 07 04.5 & +27 35 56.3 & 1.48E-08 & 9.94   & 1.89E-16 & 0.17     & 0.95       & 4.50(1.86)    & 8.9(3.56)    \\
SDSS J134855.27-0321 & 13 48 55.3 & --3 21 41.4 & 1.62E-08 & 9.89   & 1.78E-16 & 0.14     & 0.80        & 4.09(1.64)   & 7.95(3.17)   \\
SDSS J160730.33+1449 & 16 07 30.3 & +14 49 04.2 & 1.34E-08 & 9.82   & 1.44E-16 & 0.17     & 0.94    & 3.94(1.69)   & 7.88(3.2)    \\
SDSS J131706.19+2714 & 13 17 06.2 & +27 14 16.7 & 1.39E-08 & 9.92   & 1.34E-16 & 0.11     & 0.63       & 3.08(1.26)   & 6.50(2.57)    \\
SNU J13120+0641      & 13 12 04.7 & +06 41 07.6 & 1.55E-08 & 9.14   & 1.33E-16 & 0.10     & 0.57         & 2.98(1.19)   & 5.97(2.37) \\
OJ287   & 08 54 48.9 & +20 06 31  & 5.82E-09 &  10.26  & 1.11E-16   & 0.06 & 1.05 & 1.92(1.33) & 9.69(3.69)\\

\hline
\end{tabular}
\end{table*}

\begin{deluxetable}{lcc}
\tablenum{3}
\tablecaption{List of 36 SMBHB candidates with BH total mass error estimates obtained with the width of broadline spectral emissions \citep{Shen2008}, and the rate of change in their GW frequencies, in Hz per year. \label{Table:BHmass_err}}
\tablewidth{0pt}
\tablehead{
\colhead{Object Name}  & \colhead{log($M$/$M_{\odot}$)} & \colhead{$\dot{f_{GW}}$[Hz/yr]} }
\startdata
HS 1630+2355         & 9.74$\pm$0.26 & 8.98E-04   \\
SBS 0920+590         & 9.20$\pm$0.29  & 6.95E-04   \\
FBQS J081740.1+23273 & 9.38$\pm$0.28  & 1.40E-03   \\
SDSS J131706.19+2714 & 9.35$\pm$0.28 & 5.32E-04   \\
SDSS J155449.11+0842 & 9.21$\pm$0.28  & 4.87E-05   \\
SDSS J104430.25+0518 & 9.26$\pm$0.29 & 4.81E-05   \\
SDSS J143621.29+0727 & 9.22$\pm$0.28 & 3.85E-04   \\
SDSS J133127.31+1824 & 9.21$\pm$0.30 & 2.83E-04   \\
HS 0946+4845         & 9.02$\pm$0.29 & 1.65E-05   \\
SDSS J150450.16+0122 & 9.18$\pm$0.29 & 1.34E-04   \\
SDSS J144755.57+1000 & 8.95$\pm$0.29 & 6.08E-04   \\
SDSS J083349.55+2328 & 9.01$\pm$0.30 & 5.08E-03   \\
SDSS J091554.50+3529 & 9.02$\pm$0.3      & 4.90E-05   \\
SDSS J121018.66+1857 & 9.12$\pm$0.29 & 2.73E-05   \\
SDSS J082827.84+4003 & 9.13$\pm$0.29 & 1.19E-04   \\
CSO 67               & 9.02$\pm$0.29 & 1.23E-04   \\
SDSS J114438.34+2626 & 9.03$\pm$0.28 & 5.86E-05   \\
SDSS J121457.39+1320 & 8.93$\pm$0.30 & 9.96E-05   \\
SDSS J082121.88+2508 & 8.98$\pm$0.30 & 6.33E-04   \\
SDSS J152157.02+1810 & 8.86$\pm$0.31 & 5.58E-07   \\
SDSS J133654.44+1710 & 8.77$\pm$0.31 & 3.47E-05   \\
SDSS J224829.47+1444 & 8.75$\pm$0.31 & 8.85E-05   \\
SDSS J124044.49+2310 & 8.74$\pm$0.30 & 1.58E-04   \\
SDSS J221016.97+1222 & 8.93$\pm$0.29   & 7.35E-05   \\
QNZ3:54               & 8.65$\pm$0.31 & 5.30E-06   \\
SDSS J115346.39+2418 & 8.77$\pm$0.31 & 5.76E-05   \\
SDSS J115141.81+1421 & 8.70$\pm$0.31 & 6.10E-05   \\
SDSS J082716.85+4905 & 8.66$\pm$0.31 & 5.21E-07   \\
SDSS J132103.41+1237 & 8.69$\pm$0.32 & 6.78E-05   \\
SDSS J154409.61+0240 & 8.86$\pm$0.30 & 1.42E-05   \\
SDSS J170616.24+3709 & 8.55$\pm$0.31 & 4.45E-06   \\
SDSS J133807.69+3602 & 8.61$\pm$0.32 & 5.39E-06   \\
SDSS J103111.52+4919 & 8.64$\pm$0.30 & 1.61E-05   \\
SDSS J104758.34+2845 & 8.54$\pm$0.30 & 2.12E-06   \\
SDSS J130040.62+1727 & 8.62$\pm$0.30 & 2.48E-05   \\
SDSS J082926.01+1800 & 8.49$\pm$0.30 & 1.28E-05  \\
\enddata
\tablecomments{A mass uncertainty of $\sim\pm$0.3 on log-scale (shown in Column 2) results in a wide window of possible mass values for any candidate above -- the actually mass can be a factor of 2 larger ($10^{+0.3}$) or 50\% smaller ($10^{-0.3}$) than the mean BH mass.}
\end{deluxetable}

In \autoref{Fig:P_CRTS} we show the strain for the {\it CRTS} candidates, and for 37 of these we can compute the mass from SDSS spectra in order to compute the strain error bars, \autoref{Table:BHmass_err} (spectra were not immediately available for the remaining candidates). If the upper error bar touches the S/N=3 detection curve with $h_{1\sigma}$, we claim that we can constrain the binary's total mass. We also show how important the new SKA pulsars will be for PTAs -- even with short timing baselines, these new pulsars lead to significant improvements in detection prospects at mid-to-high GW frequencies. 

It is encouraging that all candidates except the weakest one, SDSS J081617.73+293639.6, in \citet{SesanaEtAl:2018} Table 1 (see also their Fig. 6) are in our Table~\ref{Table:table_of_strains}. While they rank their objects by strain from one Monte Carlo realization and compute the resulting amplitude of the gravitational wave background, we compute the the strain over $\sim10^5$ realizations and subsequently estimate the time-to-detection of these SMBHB candidates. 

The sky and polarization averaged detection curve does not however give the complete picture, since GW sensitivity is also a function of sky position and overall alignment with well-timed millisecond pulsars.
We therefore use {\tt hasasia} to generate future PTA detection sky maps, \autoref{Fig:strain_sky_ptas}.
We find that the continuous GW sensitivity is largely smoothed out over the sky by going from a single PTA (here NANOGrav) to the IPTA, as expected but never concretely shown. It is also interesting to note the residual effect the IPTA data have on SKA1's sky map, inducing a slight preference for GW observations in the northern hemisphere, while this is virtually washed out by SKA2.

Importantly, NANOGrav's most sensitive area improves by a factor of 6 in the strain, or a factor of 216 in volume, between the 11-yr data and our projections for IPTA2025. Moving from IPTA2025 to SKA1 yields a further factor of 4 improvement for the strain in 5 years, and going from SKA2 to from SKA1 yields another factor of 2 in strain sensitivity. Improvements between SKA1 and SKA2 are modest since many of the pulsars in the SKA MID and LOW surveys will likely be found by 2034 and beyond, according to \citet{KeaneEtAl:2015}.

\begin{figure*}[ht!]
 \includegraphics[width=2\columnwidth]{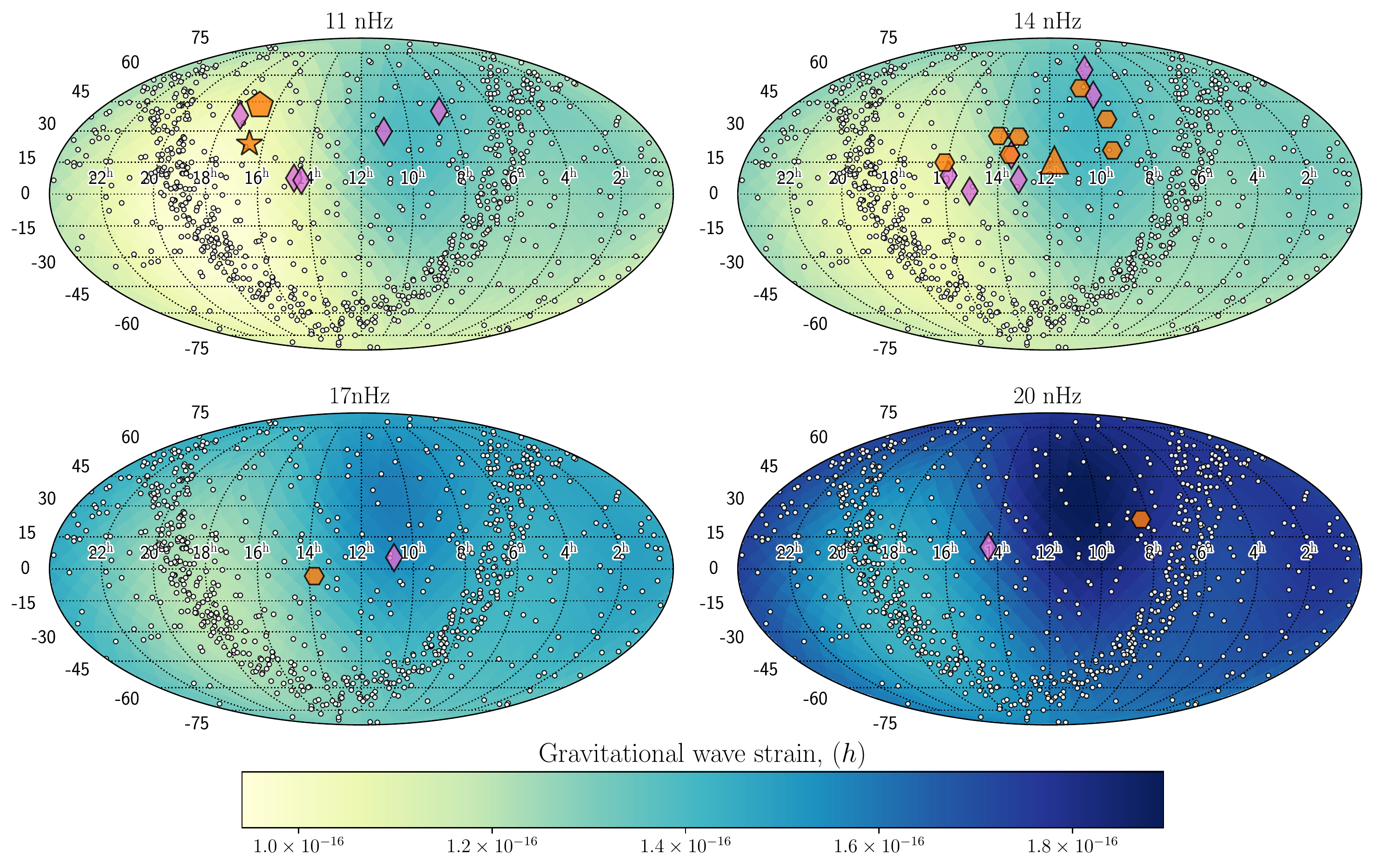}
 \caption{{\bf Minimum strain for S/N $=3$ detection for SKA1.}
 The candidates HS 1630+2355 (orange star), SDSS J164452.71+4307 (orange pentagon) and SDSS J114857.33+1600 (orange triangle) are again highlighted over the detection threshold for SKA1 at 4 different frequency bins with pulsars as white circles. The orange hexagons show the sources who's 1$\sigma$ masses will be limited in SKA1. The light purple diamonds show the sky position of the remaining {\it CRTS} sources from \autoref{Table:table_of_strains}.}
 \label{Fig:strain_sky_ska1}
\end{figure*}

In \autoref{Table:table_of_strains} we rank the selected {\it CRTS} sources from highest $1-\sigma$ strain upper limit to lowest. Using $h_{1\sigma}$ we find that 3C66B, SDSS J164452.71+4307 and HS 1630+2355 should have S/N $>3$ with IPTA2025 data if they are real binaries\footnote{Note that the strain calculations for SDSS J164452.71+4307 do not have mass errors, therefore $h_0\approx h_{1\sigma}$}. If not detected, we can start to constrain their masses, and eventually rule out these AGN as SMBHB host galaxies. 
SKA1 will constrain 12, and SKA2 will constrain 28 SMBHB candidates' masses. We also find that IPTA2025 can detect 2 sources (3 if we include the factor of 2 from \citealt{Arzoumanian2020}), SKA1 can detect 12, and SKA2 can detect 17 sources. Our findings are summarized in  \autoref{Table:table_of_strains} and \autoref{Table:table_h0}, and \autoref{Fig:strain_sky_ptas} and \autoref{Fig:strain_sky_ska1}.

We also note that SDSS J114857.33+1600's detection prospects improve by a factor of 5 from IPTA2025 to SKA1 in five years -- this is due to the large increase of low-RMS pulsars added by preliminary SKA MID and LOW surveys \citep{KeaneEtAl:2015} and timed for only up to four years. This large increase in sensitivity over a short time is also illustrated and explained by \autoref{Fig:P_CRTS}.

\begin{figure}
 \includegraphics[scale=0.6]{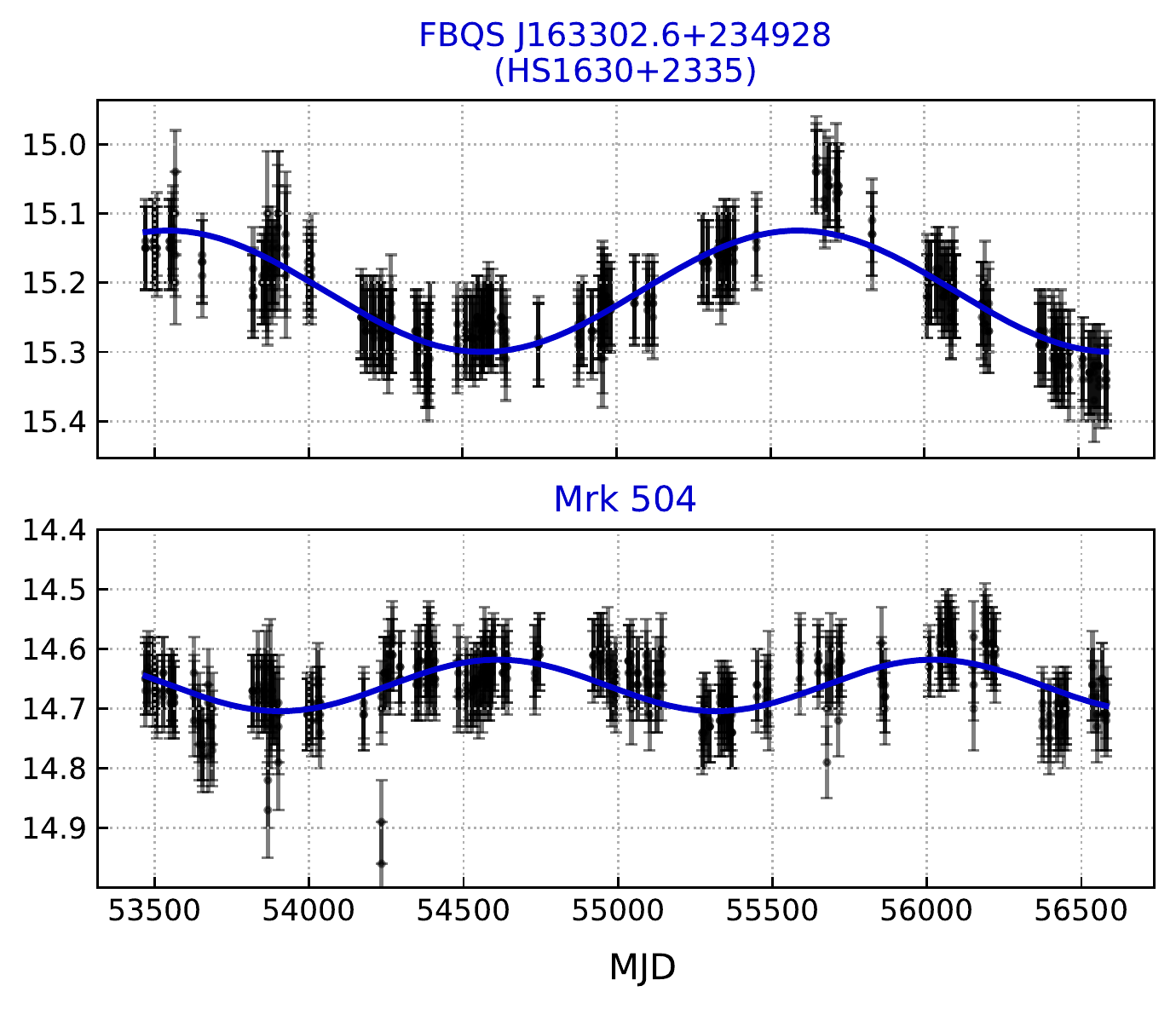}
 \caption{The optical periodic light curves of the HS1630+2335 (or FBQS J163302.6+234928 in Chandra X-ray) and Mrk 504. Their periodicity (blue) are fitted with minimized sum of squared residuals $\approx$ 6e-6 and 3e-7 respectively. The bestfit periods here are 2046 days and 1410 days, respectively. Mrk 504 is an excellent candidate to followup with PTA observation to constrain its mass, since no other mass estimate are available.}
 \label{Fig:example_LC}
\end{figure}

HS 1630+2335 and Mrk 504 show periodicity in both the optical (\autoref{Fig:example_LC}) and are not ruled out in X-ray, see \citealt{Saade2020} and references therein, making them very interesting SMBHB candidates for continued investigations. Quite serendipitously HS 1630+2335 is closely aligned with PSR J1640+2224, which is already a target in NANOGrav's high-cadence timing campaign. PSR J1640+2224's residuals have decreased by a factor of three since the first NANOGrav data release, making this pulsar a key tool to detect GWs from HS 1630+2335, or to rule it out as a true SMBHB. According to our calculations, the S/N based on $h_{1\sigma}$ of HS 1630+2335 in the NG 11-yr data is $\sim 0.9$. As a simplifying assumption, we assume all the signal comes from PSR J1640+2224 (a detection will require multiple pulsars).  Moving from the 11-yr to 12.5-yr data will quadruple the cadence, and the RMS of the residuals is half of the 11-yr value, the S/N would improve by a factor of $\sim 3$. Moreover, a dedicated search for this source in the NANOGrav 12.5-yr data would give an additional factor of 2 in sensitivity, with the albeit optimistic potential of making a detection at $h_{1\sigma}$ with S/N~$\sim5$ -- comparable to IPTA2025 sensitivity.

We show Mrk 504's EM periodicity in Figure~\ref{Fig:example_LC}, along with the photometry extracted from {\it CRTS}. It has a 10-yr data baseline, and we find its period to be 1410 days, very close to the \citep{Graham2015b} value of 1408 days. The data, ending at MJD 56580, could therefore cover about $\sim$2 binary orbital periods.
We can readily compute a mass upper limit by rearranging \autoref{eq:strain} for the chirp mass. We also compute the strain needed for an S/N=3 detection in Mrk 504's sky position using H19 (based on the NANOGrav 11-yr data) at $f_{gw}=16.4$~nHz (corresponding to an orbital period of 1410), and find this to be $h<4.4\times 10^{-15}$. Using the source's distance of 160 Mpc, we can now constrain the chirp mass to be $\mathcal{M}_c<1.4\times 10^9~M_\odot$. Assuming an equal mass ratio, $q=1$, we can limit to the total mass of the binary to be $M<3.3\times 10^9~M_\odot$. \citealt{Ho2008} uses the H--I emission line method \citep{Greene&Ho2005b} to find the BH mass of Mrk 504 (alternative name, PG 1659+294), log$(M_{BH}/M_{\odot})=6.69\pm 0.5$. Mrk 504 is also in process of reverberation mapping campaign (Lick AGN Monitoring Project) which could yield more accurate mass estimate \citep{Pancoast2019}.

Furthermore we independently compute the probability of each galaxy in {\it CRTS} hosting a SMBHB system emitting nanohertz GWs, as in \citet{Mingarelli2017}. We find that 24 high-$z$ AGN should not be able to host SMBHB systems at all, \autoref{Table:impossible-binaries}, since there has not been enough time for such binary systems to form under the assumption that they undergo a dynamical friction phase followed by a stellar hardening phase. The detection of GWs from one of these {\it unlikely} host galaxies would imply significant gas interactions and/or binary eccentricity are accelerating these high $z$ mergers -- a plausible scenario and a subject of future study -- or that these periodic light curves are not truly tracing binary activity.

\begin{deluxetable*}{lcclcccc}
\tablenum{4}
\tablecaption{Unlikely SMBHB candidates When computing the probability of each {\it CRTS} candidate hosting a SMBHB system, \autoref{eq:pmerge}, we found that 24 candidates were not viable binaries, though this may be due to the limitations of our model. For them to be viable, their parent galaxies would have had to start their mergers at $z>4$ and furthermore one would need to invoke physical processes such as high eccentricity and strong accretion-based torques for the binaries to be emitting GWs at the time of observation. \label{Table:impossible-binaries}}
\tablewidth{0pt}
\tablehead{
\colhead{Object Name}  & \colhead{z} & \colhead{log($M$)} & \colhead{strain} & \colhead{NG $\rho$,} & \colhead{IPTA $\rho$,} & \colhead{SKA1 $\rho$,} & \colhead{SKA2 $\rho$,} \\ 
\colhead{}&\colhead{} &\colhead{[$M_{\odot}$]} & \colhead{$h_0$} &  \colhead{11-yr} & \colhead{2025} & \colhead{2030} & \colhead{2034+}}
\startdata
SDSS J164452.71+430752.2 & 1.715 & 10.15 & 4.94E-16 & 0.65 & 3.67 & 13.77(6.26) & 30.82(12.73) \\
SDSS J114857.33+160023.1 & 1.224 & 9.9   & 3.03E-16 & 0.22 & 1.39 & 6.88(2.74)  & 16(6.27)     \\
HS 0926+3608             & 2.15  & 9.95  & 2.04E-16 & 0.08 & 0.74 & 4.29(1.66)  & 9.52(3.68)   \\
SDSS J092911.35+203708.5 & 1.845 & 9.92  & 2.02E-16 & 0.07 & 0.87 & 4.43(1.73)  & 10.89(4.17)  \\
SDSS J133516.17+183341.4 & 1.192 & 9.76  & 1.91E-16 & 0.18 & 1    & 4.67(1.92)  & 9.7(3.87)    \\
SDSS J140704.43+273556.6 & 2.222 & 9.94  & 1.89E-16 & 0.17 & 0.95 & 4.5(1.86)   & 8.9(3.56)    \\
SDSS J134855.27-032141.4 & 2.099 & 9.89  & 1.78E-16 & 0.14 & 0.8  & 4.09(1.64)  & 7.95(3.17)   \\
SDSS J160730.33+144904.3 & 1.8   & 9.82  & 1.45E-16 & 0.17 & 0.95 & 3.97(1.71)  & 7.94(3.22)   \\
SDSS J124119.04+203452.7 & 1.492 & 9.4   & 1.45E-16 & 0.09 & 0.63 & 3.2(1.29)   & 7.57(2.99)   \\
SDSS J131706.19+271416.7 & 2.672 & 9.92  & 1.34E-16 & 0.11 & 0.63 & 3.08(1.26)  & 6.5(2.57)    \\
QNZ3:54                   & 1.402 & 9.27  & 7.34E-17 & 0.04 & 0.19 & 0.57(0.29)  & 1.59(0.66)   \\
3C 298.0                 & 1.437 & 9.57  & 6.73E-17 & 0.08 & 0.42 & 1.8(0.74)   & 3.99(1.6)    \\
SDSS J083349.55+232809.0 & 1.155 & 9.4   & 6.73E-17 & 0.08 & 0.42 & 1.8(0.74)   & 3.99(1.6)    \\
SDSS J155647.78+181531.5 & 1.502 & 9.51  & 6.24E-17 & 0.06 & 0.35 & 1.54(0.66)  & 2.91(1.18)   \\
SDSS J094450.76+151236.9 & 2.118 & 9.61  & 6.00E-17 & 0.02 & 0.21 & 1.22(0.47)  & 2.64(1.01)   \\
SDSS J121018.66+185726.0 & 1.516 & 9.53  & 5.82E-17 & 0.04 & 0.26 & 1.32(0.53)  & 2.94(1.15)   \\
BZQJ0842+4525             & 1.408 & 9.48  & 5.02E-17 & 0.02 & 0.22 & 1.11(0.44)  & 2.82(1.09)   \\
SDSS J121457.39+132024.3 & 1.494 & 9.46  & 4.25E-17 & 0.03 & 0.2  & 0.97(0.39)  & 2.33(0.92)   \\
SDSS J082121.88+250817.5 & 1.906 & 9.53  & 4.18E-17 & 0.01 & 0.19 & 0.94(0.37)  & 2.4(0.92)    \\
SDSS J093819.25+361858.7 & 1.677 & 9.32  & 2.85E-17 & 0.01 & 0.08 & 0.51(0.2)   & 1.08(0.42)   \\
SDSS J165136.76+434741.3 & 1.604 & 9.34  & 2.45E-17 & 0.03 & 0.18 & 0.67(0.3)   & 1.47(0.61)  \\
SDSS J014350.13+141453.0 & 1.438 & 9.21  & 1.99E-17 & 0.01 & 0.1  & 0.44(0.19)  & 0.95(0.41)   \\
SDSS J123147.27+101705.3 & 1.733 & 9.2   & 1.34E-17 & 0.01 & 0.06 & 0.31(0.12)  & 0.71(0.28)   \\
SDSS J080809.56+311519.1 & 2.642 & 8.36  & 4.34E-19 & 0    & 0    & 0.01(0)     & 0.02(0.01)   \\
\enddata
\end{deluxetable*}

Importantly, we find that the existence of a GW background does not impede the detection of individual SMBHB at higher frequencies probed by {\it CRTS}, {\it PTF}, and others. Since the background is predominantly a very low-frequency GW signal, it only decreases our sensitivity by as factor of $\sim3.0$ at $\sim 11$~nHz, the lowest frequency considered here.

\section{Discussion}\label{sec:conclusion}
Our next step will be to analyze recent Sloan Digital Sky Survey (SDSS; \citealt{Liao+2020}) data and Dark Energy Survey (DES; \citealt{Chen+2020}). These further studies will be important, since with with the {\it CRTS}, SDSS, and DES surveys we may be able to find some common false-positive signals of binary activity if SMBHB candidates are completely ruled out by a lack of GW signal. Upcoming wide-field time-domain surveys such as the Rubin Observatory Legacy Survey of Space and Time (LSST; \citealt{LSST}), and ambitious spectroscopic surveys such as the Sloan Digital Sky Survey-V (SDSS-V; \citealt{SDSS-V}) and the Dark Energy Spectroscopic Instrument (DESI; \citealt{DESI}) will reveal a sky rich in potential SMBHBs systems \citep{KelleyEtAl:2019}. Indeed, this work will form an important basis for assessing the credibility of SMBHB host galaxies which can be used in the LSST and SDSS-V era, when the sky is full of SMBHB candidates.

To this end, we outline general and concrete detection strategies can be initiated to make a GW detection from a candidate SMBHB host galaxy, or to rule out a given AGN as a SMBHB host galaxy. 
To optimize our ability to detect these GWs, we propose the following strategy for the {\it CRTS} candidates which can be broadly applied to other such periodic AGN: (i) start or maintain a high-cadence observing program for pulsars closely aligned with the top candidates in \autoref{Table:table_of_strains} (or any future candidate in general). If no such pulsars exist, ones can be added via pulsar searches \citep{2019AO_white} or targeted radio follow-ups of {\it Fermi}-LAT unassociated sources (so far $\sim 20\%$ of these have been added to NANOGrav\footnote{\url{https://confluence.slac.stanford.edu/display/GLAMCOG/Public+List+of+LAT-Detected+Gamma-Ray+Pulsars} }), (ii) an increase in the amount of time spent observing these pulsars to lower their RMS white noise values, (iii) combining pulsar datasets to achieve an immediate increase in cadence, (iv) use targeted continuous GW searches to improve strain limits by a factor of at least two~\citep{Arzoumanian2020}.

These strategies have not been applied to candidates in \autoref{Table:table_of_strains} and \autoref{Table:table_h0}, and could significantly accelerate their time to detection. Moreover, a new pulsar added via e.g. {\it Fermi} is almost immediately valuable for continuous GW detection, as we showed in \autoref{Fig:P_CRTS} for the SKA. 

Some particularly interesting candidates to watch are 3C66B, which could show GW frequency evolution, HS1630 where the mass can be constrained with upcoming IPTA observations, and Mrk 504 where we were able to compute a mass upper limit, $M=3.3\times 10^9$ M$_{\odot}$. In addition to the latter being relatively nearby at 160 Mpc, this rather good mass constraint could also be due to Mrk 504's proximity to pulsar J1713+0747 -- an exceptional pulsar.

Large errors on the SMBH mass estimates from \cite{Shen2008} translate into large strain errors, since $h\propto [q/(1+q)^2]M^{5/3}$. Therefore uncertainties in $q$ will always be subdominant to uncertainties in $M$, e.g. \autoref{Fig:P_CRTS}. In fact the SMBH masses may have been overestimated by a statistical factor of 3 \citep{SesanaEtAl:2018} due to the steep slope of the BH mass function at high mass. We are therefore very interested in investigating the differences between a systematic mass overestimate of a factor of 3, versus an error in a BH host galaxy scaling relation in future work. We believe that our mass limits and the eventual detection of GWs from these systems will allow us to disentangle these competing mass errors.

Indeed, PTAs offer a unique opportunity to constrain the chirp mass $\mathcal{M}_c$ of candidate SMBHBs, which in turn provides new GW-based constraints on the underlying EM-based SMBH mass estimates, e.g. \citet{Shen2008}. However, surveys like {\it CRTS} may select quasars exhibiting red noise which only appears to be periodic over a short timespan \citep{VaughanEtAl:2016}. The detection strategies we outline, however, will allow us to identify {\it true} binary SMBHs by confirming and ruling out signals that are simply red noise, making this toolkit is complementary to previous methods, i.e.  \citealt{VaughanEtAl:2016} and \citealt{KelleyEtAl:2019}. \citet{ZhuThrane2020}'s Bayesian approach to measuring periodicity is also a promising tool to mitigate this risk and identify truly periodic signals.  

We previously discussed the limitations of Doppler boosting in identifying potentially detectable SMBHBs \citep{Dorazio2015Nature} -- such small mass ratios make detecting such binaries exceedingly difficult. However in the case of PG1302-102, we allow ourselves to be cautiously optimistic. The strain of the candidate SMBHB in this galaxy is $h\sim 9\times 10^{-18}$ at $\sim 14$~nHz \citep{Graham2015a}. Using the detection strategies we describe here, complemented by a targeted search for the additional factor of two~\citet{Arzoumanian2020} from performing a targeted GW search, may make it detectable with SKA2. Additionally, \citet{DOrazio2015} and \citet{Noble2021} use hydrodynamical simulations to demonstrate that SMBH binaries with more equal mass ratios, $q\gtrsim 0.3$, which is the adopted assumption in our calculation of strains ($0.25<q<1$), show EM periodicity at orbital periods a factor of $\sim$few shorter due to an overdensity in the circumbinary disk. This would generally result in larger strains than in our predictions.

Some 24 high-mass, high-$z$ AGN in {\it CRTS} are not viable candidates according to our SMBHB evolution models: the host galaxies would have had to merge at $z>4$ to give rise to these binaries, and our current methodologies only extend to $z=4$ due to limitations in Illustris \citep{Rodriguez-Gomez2015}. Moreover, the SMBH occupation fraction for the parent galaxies which formed the current binaries would have to be 1.0 at $z>4$, which would also be an interesting result. Our models do not, however, include gas and potential binary eccentricity, both of which would make the binaries evolve more quickly. A GW detection from one of these 24 host galaxies, such as SDSS J164452+4307 (detectable by IPTA in 2025, \autoref{Table:table_h0}), could therefore inform SMBHB evolution models, and may help in understanding the origins of SMBHs seeds \citep{VolonteriEtAl:09, TanakaHaiman:09} and their occupation fraction.

Importantly, we found that the presence of a stochastic GW background does not impede our detection prospects. In fact, if these AGN really are SMBHB systems, they may induce some anisotropy in the GW background~\citep{MingarelliEtAl:2013, Mingarelli2017}. Detection of anisotropy will likely follow that of the isotropic GW background in the next few years. 

So far, no CGWs have been detected at any frequency from any compact object. Long-lived CGWs  are currently detectable with PTAs, and may eventually be complemented by astrometric GW detection \citep{MooreEtAl:2017}. Given that SMBHBs in the PTA band will stay in band for tens of millions of years, the current periodic light curves in {\it CRTS}, {\it PTF}, and {\it PAN-STARRS1} offer us a unique opportunity to study SMBHB host galaxies and their EM emissions. Identifying these EM emissions will be invaluable information for the Laser Interferometer Space Antenna (LISA; \citealt{lisaWhitepaper}), where the angular resolution is relatively poor, and the signals will only last weeks to months. Understanding the expected EM emissions from SMBHB host galaxies is thus important groundwork to lay for both PTAs and LISA in this new era of multimessenger astrophysics.

\section{Acknowledgements}
The authors thank William Garnier for an updated SKA1 timeline, and Zolt\'an Haiman, Maria Charisi, Michael Lam, M. Lynne Saade, Alberto Sesana, Yacine Ali-Ha\"imoud, Thomas McCausland, and Sean Oh for useful discussions.
The Flatiron Institute is supported by the Simons Foundation. The NANOGrav project receives support from National Science Foundation (NSF) PIRE program award number 0968296 and NSF Physics Frontier Center award number 1430284. CX was supported by NASA ADAP grant NNX17AL82G.

\software{
    Astropy \citep{Astropy2013, Astropy2018},
    Hasasia \citep{HazbounHas:2019},
    Healpy \citep{Healpy2019}, 
    Matplotlib \citep{Hunter2007}, 
    Numpy \citep{Numpy2011},
    Pandas \citep{Pandas2010}, 
    SciPy \citep{SciPy2019},
    NanohertzGWs \citep{MingarelliGWs}
}

\end{document}